\documentclass[12pt]{article}
\usepackage[dvips]{graphicx}
\usepackage{amsmath,amsthm,amssymb}

\usepackage{subfigure}

\newcommand{\bbZ}{\mathbb{Z}}

\theoremstyle{definition}

\begin{document}
\baselineskip 21pt

\title{Retrial Queueing Models: A Survey on Theory and Applications}


\author{
        Tuan Phung-Duc \\
        Faculty of Engineering, Information and Systems \\  
		University of Tsukuba \\
		1-1-1 Tennodai, Tsukuba, Ibaraki 305-8573, Japan \\
		Email: tuan@sk.tsukuba.ac.jp
}

\date{}

\maketitle

\begin{abstract}
Retrial phenomenon naturally arises in various systems such as call centers, cellular networks and random access protocols in local area networks. This paper gives a comprehensive survey on theory and applications of retrial queues in these systems. We investigate the state of the art of the theoretical researches including exact solutions, stability, asymptotic analyses and multidimensional models. We present an overview on retrial models arising from real world applications. Some open problems and promising research directions are also discussed.
\end{abstract}



\section{Introduction}
%
%
%
%
The loss models (including Erlang loss model) assume that an arriving customer that sees the service area being fully 
occupied is blocked and is lost forever. On the other hand, in models with an infinite waiting capacity, 
a customer waits until being served. However, there are various situations in our everyday life and in various systems where blocked customers are not willing to wait and they temporarily leave the service facility for a while but try again after some random time. A blocked customer is said to be in a virtual waiting room called {\it orbit} before retrying to occupy a server again. These situations are modeled as retrial queues.

\begin{figure}[bthp]
\begin{center}
\includegraphics[scale=0.3]{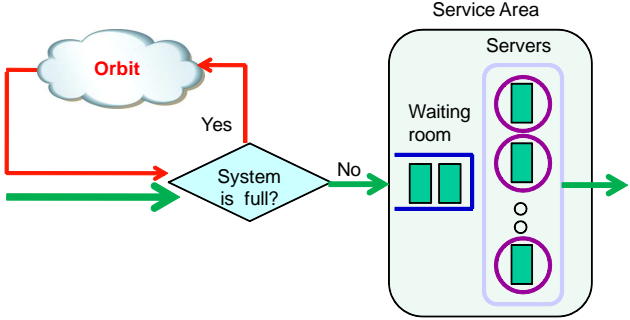} \\
\caption{A general multiserver retrial queue.}
\label{fig:multiserver_retrial_queueing_model}
\end{center}
\end{figure}

For example, in a call center, if a customer makes a phone call when all the agents are busy, the customer will try to 
make a phone call again after some random time. In computer networks, if a packet is lost, the packet 
may be retransmitted at a later time by a retransmission mechanism such as the TCP (Transmission Control Protocol)~\cite{Avrachenkov1,Avrachenkov2}. 
In these applications, the orbit is virtual and cannot be observed. Figure~\ref{fig:multiserver_retrial_queueing_model} 
represents a general multiserver retrial queue. 

In many applications, the customers in the orbit act independently of each other, thus the retrial rate depends on the 
number of customers in the orbit. As a result, the underlying Markov chains of retrial queues have a spatially non-homogeneous structure. 
Due to the spatial non-homogeneity of the underlying Markov chains, the analysis of retrial queues is more complex and challenging than that of standard queues. For an extensive comparison of standard and retrial queueing systems, the readers are referred to 
the paper of Artalejo and Falin~\cite{artalejo_falin02}. Even for the M/M/$c$/$c$ retrial queue, where retrial interval 
of customers follows an exponential distribution, analytical solutions are obtained in only a few special cases~\cite{artalejo_falin02,Fali97}. 

\subsection*{Classical retrial policy}
In a retrial queue with the classical retrial policy, each blocked customer 
stays in the orbit for an exponentially distributed time independently of other customers. 
As a result, the retrial rate is proportional to the number of customers in the orbit. 
The classical retrial policy naturally arises from applications such as call center and telephone 
exchange system, where customers in the orbit act independently because a retrial customer 
cannot observe the behavior of the others. Retrial queues with the classical retrial policy  
have attracted many researchers since they naturally arise from various applications with 
random access. Under some Markovian settings of the service time distribution and the arrival process, the 
underlying stochastic process is a level-dependent quasi-birth-and-death (LDQBD) process where the level is the number of customers in the orbit and the phase represents the states of the servers. For general LDQBD processes, some general numerical algorithms are available~\cite{bright95,phung3}. The sparsity of the block matrices of the LDQBD processes of retrial queues could be used to develop efficient computational algorithm~\cite{phung4}.


%
%
\subsection*{Constant retrial policy}
There are several applications in communication networks, where the retrial of customers is controlled. It means that the retrial rate may not depend on the number of customers in the orbit. For example, Choi et al.~\cite{BDChoi92} study the stability of the CSMA/CD (Carrier Sense Multiple Access with Collision Detection) protocol, by a retrial queue with a constant retrial rate.  
Avrachenkov and Yechiali~\cite{Avrachenkov1,Avrachenkov2} use a retrial queueing network with 
constant retrial rate to model TCP traffic. Constant retrial rate could be interpreted by the so-called ``calling for blocked customers". When the server is idle, it calls blocked customers one by one. The time for the server to pick up a blocked customer could be interpreted as the retrial time.  

Artalejo et al.~\cite{Artalejo_gomez_neuts} formulate a multiserver retrial queue (M/M/$c$/$c$) with constant retrial rate by a level-independent QBD process which could be analyzed efficiently using matrix analytic methods invented by Neuts~\cite{Latouche99,Neut94}. As in the classical retrial rate, the block matrices of the level-independent QBD process are also sparse leading to efficient algorithms for the stationary distribution. These algorithms are discussed by Artalejo et al.~\cite{Artalejo_gomez_neuts} using a matrix analytic method and by Do et al.~\cite{} using the spectral expansion method~\cite{tien_chakka10,tien_10}.

The structure and aim of this chapter are as follows. First in Section~\ref{applications}, we provide a comprehensive review of retrial queues in real world applications. Second, in Section~\ref{theory}, we present the main results on the analysis of retrial queueing models. The aim of this survey is to provide a guide for researches who want to enter and deepen the understanding of the field of retrial queues. To this end, we point out some open problems and promising research topics.

\section{Retrial Queues in Applications}\label{applications}
In this section, we present several retrial queueing models arising from real world applications. 
\subsection{Call Centers}
A call center is important for a company because it provides a channel for customers to contact the company. 
In a call center, agents are the people who answer the calls from the customers. When a customer makes a phone call, 
if there is an idle call agent, the customer is immediately answered by the call agent. If all the agents are busy, 
the customer may hear some massage such as ``the system is busy at the moment, please wait for a moment". 
At this moment, the customer either hangs up the phone immediately or continues to hear the message. 
In the former case, the customer may try again after some time. Customers who decide to wait for a free 
call agent may renege if the waiting time is too long. These customers may also make a phone call later. 

\begin{figure}[bthp]
\begin{center}
\includegraphics[scale=0.3]{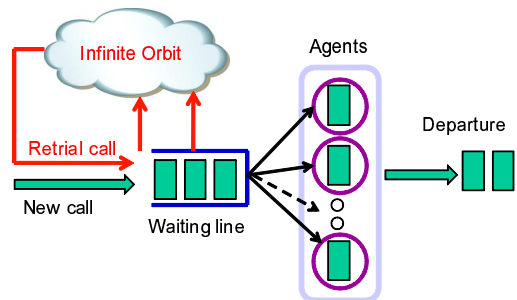} \\
\caption{A retrial queueing model for call centers.}
\label{fig:call_center_model}
\end{center}
\end{figure}

One of the most important performance measures for a call center is the blocking probability that a customer 
cannot find a free call agent upon arrival. From the customer's point of view, a low blocking probability is desirable. 
In order to keep the blocking probability to be low, a simple solution is to increase the number of agents. 
However, from a management point of view we need to minimize the number of agents, due to the fact that 
the cost of a call center is mainly the human cost~\cite{Stolletz03}. In order to achieve the customer satisfaction under some 
constraint on the cost, we need some mathematical model to express the trade-off between the customer 
satisfaction and the human cost. A queueing model is one of the most appropriate mathematical models for 
the design of call centers. In addition, in order to capture the retrial phenomenon as presented above, a 
retrial queueing model is expected to be more appropriate than the corresponding standard queueing 
model~\cite{Gans_koole_mand02,Koole_mandelbaum02,Stolletz03,salah_aguir}.  
See Figure~\ref{fig:call_center_model} for a simple retrial queueing model for call centers. 
For a detailed explanation on call centers, we refer to the book by Stolletz~\cite{Stolletz03}. 
Numerical results by Phung-Duc et al.~\cite{phung4} show that approximating retrial flow by a Poisson process leads to a large error under the fast retrial regime. This is an evidence for modelling call centers by retrial queues.

Phung-Duc and Kawanishi~\cite{Phung-Duc_Kawa14a} consider two-way communication retrial queueing systems as the models for blended call centers. Queueing models without retrials for blended call centers are proposed and analyzed in~\cite{Bulai03}. Two-way communication retrial queues are studied by Artalejo and Phung-Duc in~\cite{Artalejo12,Artalejo13} where some analytical results such as the stability condition and generating functions are obtained. The main contribution in Phung-Duc and Kawanishi~\cite{Phung-Duc_Kawa14a} is to propose an efficient computational algorithm for multiserver retrial queues with distinct distributions of incoming and outgoing calls. The algorithm~\cite{Phung-Duc_Kawa14a} could be considered as a matrix version of the one by Phung-Duc et al.~\cite{phung4}. In two-way communication queueing systems, a server not only receives incoming calls but also makes outgoing calls to outside in its idle time. This is the situation in blended call centers where the operators may make outgoing calls to the customers for some marketing purposes etc.

Furthermore, Phung-Duc and Kawanishi~\cite{Phung-Duc_Kawa14} analyze a fairly general and practical retrial model for inbound call centers with after-call work and abandonment. After-call work is a typical task in call centers where after a conversation with a customer, the operator should do some after-call work for that customer after the customer departs from the system. It means that a call line is released for a newly arrived customer. In~\cite{Phung-Duc_Kawa14}, the effects of retrials by blocked and abandoned customers on the waiting time distribution are investigated.  

\subsection{Cellular Communication Networks}
In a cellular network, the service area is divided into cells. 
Users (mobile stations) in each cell are served by a base station with 
a limited number of channels. Therefore, only a limited number of users 
can communicate at the same time. A base station serves two types of calls: 
fresh calls and handover calls. A fresh call is made by a user that stays inside 
the current cell and a handover call is made by a user that has been traveling 
from an adjacent cell into the current cell. Because a handover call has been communicating 
by a channel in an adjacent cell, the call should be assigned 
a channel upon its arrival into the current cell as soon as possible for a 
continuous communication. Therefore, a handover call should be given a priority over a fresh call. 
\begin{figure}[bthp]
\begin{center}
\includegraphics[scale=0.3]{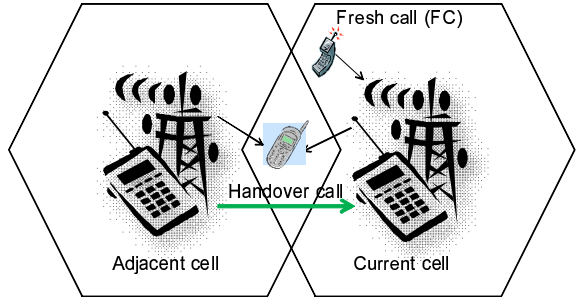} \\
\caption{Two adjacent cells in cellular networks.}
\label{fig:cellular_two_cells}
\end{center}
\end{figure}

There are several channel assignment policies that provide some priority 
for handover calls over fresh calls \cite{tien_10,Tran-Gia}. 
In a guard channel policy, a number of channels are reserved for handover 
calls. The rest of channels are equally shared by both fresh and handover calls. 
In a fractional guard channel policy, fresh calls are accepted with a probability 
depending on the number of currently occupied channels, while handover calls 
are accepted as long as an idle channel is available \cite{ramjee_towsley_nagarajan}. 
Both fresh calls and handover calls may be blocked due to the limit on the number of channels. 
In modern cellular communication systems (e.g. mobile phone system), if a call is blocked, a redial 
can be made easily, for example just by pushing only one button. In some applications, blocked 
calls are automatically redialed. Thus, the consideration of retrial calls is very important while 
designing these systems~\cite{Tran-Gia}. 

\begin{figure}[bthp]
\begin{center}
\includegraphics[scale=0.3]{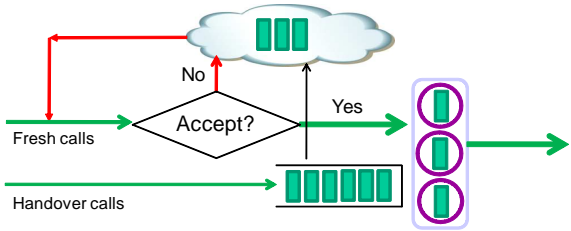} \\
\caption{A model of cellular networks with overlapping cells.}
\label{fig:cellular_networks_FGCs}
\end{center}
\end{figure}

Figure~\ref{fig:cellular_two_cells} represents two adjacent cells in a cellular networks.
In Figure~\ref{fig:cellular_networks_FGCs}, a buffer for handover calls represents overlap areas of the current 
cell with the adjacent cells. Handover calls in the overlap areas can receive signals from both the adjacent cell and 
the current cell. Thus, if a channel in the current cell is not yet available, the handover call can continue to  
communicate using the occupying channel in the adjacent cell. However, the handover call is terminated if it 
exceeds an overlap area but no idle channel in the targeting cell is available. In this case, the handover call 
may attempt again after some random time as a fresh call.

In addition to the pure Markovian models presented in \cite{tien_10,Tran-Gia}, more general models with correlated arrival processes such as Markovian Arrival Processes (MAP) and Batch Markovian Arrival Processes (BMAP) and phase type service time distributions are investigated in \cite{Alfa02,Choi99_map,Cskim10b,CSKim14}. A model under random environment is presented in \cite{artalejo10a}. In all these models, the performance measures are directly calculated from the stationary probabilities. Economou and Lopez-Herrero~\cite{Economou09} provide some more sophisticated performance measures such as waiting time distribution, idle time of guard channels etc. 

%
%
%
%
\subsection{Local Area Networks}
In a local area network (LAN), multiple nodes share a physical link (channel) in order to transmit their data (packets). 
Assuming that multiple nodes send their packets at the same time, a collision may occur and 
all the packets will be destroyed. 


Because nodes in a LAN are located closely to each other, propagation delays are much shorter than data transmission time. Therefore nodes can obtain useful information about the channel by sensing the presence of signals on the channel. Taking this property into account, CSMA (Carrier Sense Multiple Access) protocols have been developed. The fundamental idea of a CSMA protocol is that it senses the channel 
before transmitting data. In a 1-persistent CSMA protocol, when a node is ready to send its data, 
the node checks if the channel is busy. If the channel is busy, then the node continues to sense the 
link until the channel is idle and then immediately sends a frame. In case of collision, the node  
waits for a random period of time and tries to transmit again. 
The problem of the 1-persistent CSMA protocol is that in a highly loaded condition, there may be 
several nodes waiting for the availability of the channel and they send data at the same time when 
the channel becomes idle. Therefore, collisions occur with a high probability. 

Another protocol is non-persistent CSMA, in which if a node is ready to send data, it senses the channel 
and transmits data immediately if the link is available, otherwise the node waits for a random time and tries 
to retransmit again. A $p$-persistent CSMA protocol is considered to be a hybrid of 1-persistent CSMA and 
non-persistent CSMA protocols. In a $p$-persistent CSMA, if the channel is idle, the node transmits a frame
with probability $p$ and delays its transmission for one unit time (channel propagation delay) with probability $1-p$. 
If the channel is busy, the node continues to sense until the channel becomes idle and repeat the same procedure to send its frame. 

Because the collisions in CSMA protocols cannot be completely avoided, the retrial phenomenon occurs in local 
area networks, e.g. wireless networks. Therefore, retrial queueing models are considered to be more appropriate than standard 
queueing models in modelling and performance analysis of these protocols~\cite{BDChoi92,Gomez10}. 
The readers are referred to the book by Kurose and Ross for a detailed explanation on protocols for local 
area networks~\cite{kurose10}. Because the behaviors of nodes in radio networks depend on each other, analysis of an 
exact model representing the states of all nodes is very challenging. Assuming that the behaviors of nodes are independent, 
Bianchi~\cite{Bianchi2000} demonstrates a simple analysis of this complex system.  Recently, Fiems and Phung-Duc~\cite{Fiems16} propose a light traffic analysis by a series expansion method for a retrial model which represents all the states of nodes concurrently. The accuracy of the model is validated by simulations. The series expansion method~\cite{Fiems16} might be useful for other retrial models for local area networks with random access protocols.

\subsection{Retrial Queues for Cognitive Networks}
Recently, cognitive radio networks are extensively studied. The spectrum for wireless networks is physically limited. On the other hand, the amount of traffic by smart phones and other devices vastly increases day by day. However, most of bandwidth is granted to licensed users (primary users). The bandwidth is not always used by primary users. The idea of cognitive networks is to provide the opportunity for secondary users to use this bandwidth when it is not used by the primary users. Secondary users are interrupted upon the arrivals of primary users. Queueing analysis of cognitive networks has been presented by Konishi et al.~\cite{Konishi13} using a multiserver priority model without buffer where interrupted secondary users are lost. In practice, interrupted users may retry again. From this point of view, appropriate retrial queueing models might be useful for cognitive systems. 
Wang et al.~\cite{Wang-Wang16,Wang-Li16} use some simple M/M/1/1 retrial queueing models to study strategic behavior of secondary users in cognitive networks. Dudin et al.~\cite{Dudin16} propose a multiclass retrial queueing model for cognitive systems. 
Salameh et al.~\cite{Turck16}study retrial queueing models taking into account the sensing time of interrupted secondary users. Because cognitive radio is promising solution for the insufficiency of wireless spectrum~\cite{Konishi13}, new retrial models for various scenarios and technologies in cognitive systems are promising topics for future researches.

\subsection{Retrial Queues in other Applications}
Beside concrete applications mentioned in previous subsections, retrial queues are also ubiquitous in other applications. Retrial queues for optical networks are presented in~\cite{phung2_obs,Fiems-dorsman-rogiest,boxma-jacques}.
Furthermore, some emerging technology such as cloud computing, one can also find various situations where retrial queues are applicable. For example, single server retrial queues with setup time are proposed for power-saving servers~\cite{Phung_qtna2015a,Phung_qtna_jimo}. Phung-Duc and Kawanishi~\cite{Phung-Duc_Kawanishi16} investigate the impact of retrial phenomenon on power saving data centers by an M/M/$c$/$c$ retrial queue with setup time. In cloud systems, the computing unit and the storage unit may be separated leading to the transmission time between them. A retrial model for such a situation is initiated in \cite{Phung_qtna2015c}. Furthermore, in cloud systems, the capacity of the server can be scaled according to the workload in the system. Taking this into account, Phung-Duc et al.~\cite{Phung_qtna2015b,Phung_qtna_jimoa} propose single server retrial queueing models with speed scaling and setup time where the speed of the server is proportional to the number of jobs in the system.

\section{Models and Methodologies}\label{theory}
In the analysis of retrial queueing models, we need to keep track not only of the state of the servers but also the number of customers in the orbit. Under the assumption that retrial customers behave independently, the retrial flow by repeated customers makes the underlying stochastic process non-homogeneous. As a result, the analysis of retrial queues is more difficult than the corresponding models with infinite waiting room. Indeed, the model with infinite waiting room could be obtained from the corresponding retrial models by taking the limit as the retrial time tends to zero. In this section, we show methods and analytical results for some major retrial queueing models.

\subsection{M/G/1/1 Retrial Queues and Their Variants}
In this model, there is only one server and there is not a waiting room before the server. Customers arrive at the system according to a Poisson process and the service time for a customer is arbitrarily and identically distributed. An arriving customer immediately occupies the server if it is idle. Otherwise, the customer will join the orbit from which he will retry again in an exponentially distributed time with positive mean. This model could be analyzed by either embedded Markov chain or by complementary variable method as for the original M/G/1 model without retrials.

Let $\pi_{i,n}$ denote the stationary probability that the server is at state $i$ ($i = 0$ if the server is idle and $i=1$ if the server is busy) and there are $n$ customers in the orbit. Furthermore, let $\Pi_i (z) = \sum_{n=0}^\infty \pi_{i,n} z^n$ ($i=0,1$). For this model, the generating functions, i.e., $\Pi_i (z)$ ($i=0,1$) of the number of customers in the orbit can be obtained in an integral form~\cite{Fali97}. Furthermore, $\pi_{0,0}$ is given in an integral form and other probabilities $\pi_{i,j}$ are recursively computed. If the service time is exponentially distributed, the integral forms of $\Pi_i (z)$ ($i=0,1$) become explicit.  Furthermore, under the light-tailed assumption of the service time, asymptotic formula for the joint queue length distribution is known. In particular, $\pi_{0,n} \asymp C_0  n^{a-1} \sigma^{-n}$ and $\pi_{1,n} \asymp C_1  n^{a} \sigma^{-n}$ as $n \to \infty$~\cite{Kim_Kim_Ko} for some positive constants $C_0$, $C_1$, $a$ and $\sigma > 1$. 

Artalejo and Phung-Duc~\cite{Artalejo13} study M/G/1/1 retrial queue with two-way communication where the server makes an outgoing call in an exponentially distributed idle time with mean $1/\alpha$. In~\cite{Artalejo13}, incoming calls and outgoing calls follow two distinct arbitrary distributions. Under light-tailed assumption of the service time distributions of incoming and outgoing calls, asymptotic analysis of the joint queue length is also presented in~\cite{Artalejo13} using a simpler method in comparison with that of Kim et al.~\cite{Kim_Kim_Ko} for the M/G/1/1 retrial queue without outgoing calls. Under some heavy tailed assumptions of the service times, the queue length asymptotics is presented in Shang et al.~\cite{Shang06} for the M/G/1/1 model and by Yamamuro~\cite{Yamamuro} for $M^X$/G/1/1 retrial queue. 

Heavy traffic asymptotics is presented in Falin~\cite{Fali97}. In particular, when the traffic intensity is close to one, the scaled queue length distribution tends to Gamma distribution. Furthermore, when the retrial rate is extremely low, the distribution of the scaled number of customers in the orbit tends to Gaussian distribution. 
Sakurai and Phung-Duc~\cite{Sakurai16} study heavy traffic analysis for M/G/1/1 retrial queue with two-way communication. In addition to the two heavy traffic regimes: i) traffic intensity is close to 1 and ii) extremely slow retrial rate, the authors also study the regime iii) where length of an outgoing call is extremely long. Sakurai and Phung-Duc prove that in regime iii), the distribution for the scaled number of customers in the orbit tends to Gaussian distribution. 
We refer to recent work of Nazarov et al.~\cite{Nazarov16} and Fedorova~\cite{Fedorova15} for recent development on heavy traffic analysis of retrial queues without and with random environment.

An extension of the M/G/1/1 retrial queues with nonpersistent customers is proposed by Yang et al.~\cite{yang90}. In this model, a blocked fresh customer joins the orbit with probability $p$ while a blocked retrial customer joins the orbit with probability $q$, respectively. With the complementary probability, they abandon joining the orbit. In the case $q=1$, the analysis is almost the same as that of the basic model with $p=q=1$. However, the analysis for the case $q < 1$ is essentially different from that of the case $q=1$. In particular, although the equations for the partial generating functions of the joint queueing length distribution are obtained, the solution for these equations is not obtained. It is shown that all the quantities interest such as the joint stationary joint queue length distribution and the factorial moments are expressed in terms of the utilization of the server which is unknown. A numerical algorithm is then developed to determine the utilization of the server. However, analytical expression for the utilization remains an open problem. 
\subsection{Multiple Server Models}
While the single server model is relatively tractable in comparison with that without retrials, multiserver retrial queues are much more difficult than those without retrials. The basic M/M/$c$/$c$ retrial queues with classical retrial rate (i.e. exponential retrial intervals) has been studied by Cohen~\cite{Cohen57}. In this system customers arrive at the system according to a Poisson process with rate $\lambda$, the service time of a customer is exponentially distributed with mean $1/\nu$ and the retrial intervals are exponentially distributed with mean $1/\mu$.
  
In this model, let $\pi_{i,j}$ denote the joint stationary probability that there are $i$ ($i=0,1,\dots,c-1,c$) busy servers and $j$ customers ($j \in \bbZ_+$) in the orbit. Furthermore, let $\Pi_i(z) = \sum_{j=0}^\infty \pi_{i,j} z^j$ ($i=0,1,\dots,c$). Explicit expressions for $\pi_{i,j}$ and $\Pi_i(z)$ are obtained for the case $c=1$. For $c=2$, $\pi_{i,j}$ and $\Pi_i(z)$ are expressed in terms of hypergeometric functions~\cite{Hans87}. For the case $c \geq 3$, it is challenging to obtain analytical solutions for $\pi_{i,j}$ and $\Pi_i (z)$. Pearce~\cite{Pear89} constructs a solution to the balance equations in terms of generalized continued fraction. Phung-Duc et al.~\cite{phung1} prove for the case $c=3,4$ that $\pi_{i,j}$ is expressed in terms of the minimal solution of a three-term recurrence relation and thus in terms of continued fractions. 

Phung-Duc et al.~\cite{phung2} extend the analysis to M/M/$c$/$c$ ($c \leq 4$) retrial queues with nonpersistent customers where a blocked fresh customer and a blocked retrial customer joins the orbit with probability $p$ and $q$, respectively. As shown in~\cite{phung2}, the solution for the model $q=1$ is almost the same as that of the basic model with $p=q=1$, while the solution for the case $q<1$ exhibits a different structure. Furthermore, an accurate algorithm is developed to calculate these continued fractions with pre-specified accuracy~\cite{phung2}.

Tail asymptotic analysis for the queue length distribution of M/M/$c$/$c$ retrial queues has been extensively investigated.
Liu and Zhao~\cite{Binliu10} show that $\pi_{c-i,j}$ is of the order of $j^{a-i} \rho^j$ as $j \to \infty$ for some constant $a$ where $\rho = \lambda/(c\nu) <1$, using a matrix analytic method. 
The analysis of Liu and Zhao~\cite{Binliu10} is based on the series expansion (up to second order) of the rate matrix $R_n$ of the underlying level-dependent QBD process in terms of $1/n$. The series expansion is extended to any order by Phung-Duc~\cite{Phung-Duc15} for M/M/$c$/$c$ models with two types of non-persistent customers. It should be noted only the last row of $R_n$ is non-zero in these models. Kajiwara and Phung-Duc~\cite{Phung-Duc-Kaji16} further extend the analysis of Phung-Duc~\cite{Phung-Duc15} to an M/M/$c$/$c$ retrial model with one guard channel for priority and retrial customers where the last two rows of the rate matrix $R_n$ are non-zero. Kim et al.~\cite{Kim-Kim-Kim12,Kim-Kim12} refine the results of~\cite{Binliu10} by a generating function approach.

A simple and accurate fixed point approximation is proposed by Cohen~\cite{Cohen57} for the case where the retrial rate is relatively small in comparison with the service rate. In such a situation, the total arrival flow by both fresh customers and retrial ones is approximated by a Poisson process whose rate is the solution of a fixed point equation (See e.g. Falin~\cite{Fali97} for details). 
Let $ \lambda + r$ denote the arrival rate of the approximated Poisson process, where $r$ is the additional arrival rate due to retrial customers. Let $B(\lambda,c)$ denote the blocking probability of the corresponding Erlang-loss system without retrials where $\nu=1$. The additional arrival rate $r$ is the solution of   
\begin{equation}
\label{cohen_eq}
    r = (\lambda + r) B(\lambda + r,c).
\end{equation}   
The main reason  for this approximation is that under low retrial regime, the retrial flow is likely to form a Poisson process. 
Recently, the heavy traffic analysis (also known as Halfin-Whitt regime) for equation (\ref{cohen_eq}) is presented by Avram et al.~\cite{Avram13}.

For a numerical computation of the joint stationary distribution, we need somehow to truncate the orbit at some truncation point. The simplest truncation method is limiting the number of customers in the orbit to $N$. Thus, a blocked customer that sees $N$ customers in the orbit is lost. Another truncation method is to modify the structure of the original Markov chain after the truncation point. The modification makes Markov chain analytically tractable. This is referred to as generalized truncation in the literature~\cite{Fali97,Neut90,Anisimov02,artalejo02}. The idea in~\cite{Fali97,artalejo02} is to disregard the states (after the truncation level (orbit size)) having small probability mass.

More general models with MAP or BMAP arrivals and phase-type service time are also investigated~\cite{Breuer02,Klimenok_Dudin06,Cskim10b,Cskim10} by means of the so-called quasi-Toeplitz Markov chains~\cite{Klimenok_Dudin06}.

\subsection{Stability Conditions}
The stability condition of the basic M/M/$c$/$c$ retrial queue is simply given by $\lambda < c\nu$ or the offered load ($\lambda/\nu$) is less than the number of servers. The proof of this result is based on an appropriate Lyapunov function of a linear form $f(i,j) = ai + j$, where $i$ is the number of busy servers and $j$ is the number of customers in the orbit~\cite{Fali97}. Artalejo and Phung-Duc~\cite{Artalejo12} derived the necessary and sufficient condition for M/M/$c$/$c$ retrial queues with two-way communication where an idle server may make an outgoing call after some exponentially distributed idle time with mean $1/\alpha$. In~\cite{Artalejo12}, incoming calls and outgoing calls follow two distinct exponential distributions. It turns out that the stability condition of the M/M/$c$/$c$ retrial queues with two-way communication coincides with that of the corresponding model without outgoing calls, i.e., $\lambda < c\nu$. Phung-Duc and Dragieva~\cite{Phung-Duc-Dragieva} obtain the stability condition for a multiserver retrial queue with interaction between servers and orbit where not only customers retry but also servers call for customers from the orbit. The Lyapunov function for the models in \cite{Artalejo12,Phung-Duc-Dragieva} is of the form $f(i,j,k) = ai + bj + k$ where $i$ and $j$ are variables representing the states of the servers and $k$ is the number of customers in the orbit.

For retrial models whose LDQBD process has complex phase structure, it is efficient to use the Lyapunov function proposed by Diamond and Alfa~\cite{Diamond98}. This approach has been used to show the stability condition for MAP arrival models, model with after call work~\cite{Phung-Duc_Kawanishi11,Phung-Duc_Kawa14} and model with setup time~\cite{Phung-Duc_Kawanishi16}.

In a recent paper, Shin~\cite{Shin14} proves that the stability condition for the multiclass M/M/$c$/$c$ retrial queue is that the total offered load of all classes is less than the number of servers. The proof of Shin~\cite{Shin14} is based on a Lyapunov function that is a linear combination of the numbers of customers in the orbits of all classes and the states of the servers.
The stability condition concerns when there are a very large number of customers in the orbit. Thus, the time that a retrial customer reaches an idle server is almost zero in such a situation. This is an intuitive observation that the stability condition of retrial queues coincides with that of corresponding models with infinite buffer. Recently, Dayar and Can Orhan~\cite{Dayar16} prove the stability condition for multiclass MAP/PH/$c$ retrial queues with  cyclic PH retrial times. The proof in~\cite{Dayar16} is based on a Lyapunov function which is also a linear function of the numbers of customers in the orbit. It should be noted that the Lyapunov function approach is applied for Markovian models only.  

In a fairly general class of retrial queues (both single and multiple class models) with classical retrial rate, the coincidence in the stability conditions between retrials models and non-retrial models is confirmed by Morozov et al.~\cite{Morozov07,Morozov16c}  for more general non-Markovian models (i.e., arbitrary renewal arrivals and arbitrary service time) using the regenerative approach. The regenerative approach of Morozov et al. is also used to prove other models with constant retrial rate~\cite{Morozov14,Morozov16}. The result in~\cite{Morozov16c} generalizes that of Shin~\cite{Shin14}.

\subsection{Multiclass Models}
\subsubsection{Classical retrial policy}
We consider the multiclass M/G/1/1 retrial model with classical retrial rate where $m$ classes of customers arrive at the server according to $m$ distinct Poisson processes with rate $\lambda_1, \lambda_2, \dots, \lambda_m$. The service times of $m$ classes of customers follow $m$ distinct arbitrary distributions. A blocked customer of class $k$, joins the $k$-th orbit and retries to enter the server after some exponentially distributed time with mean $1/\mu_k$ ($k=1,2,\dots,m$). For this model and its variants the stability conditions are available and the means number of customers (also mean waiting time) in the orbit for each class are obtained~\cite{Fali97,Fali88,Kulkarni86}. Grishechkin~\cite{Grishechkin92} studies single server retrial queue with structured batch arrivals and investigates some heavy traffic limits for the queue length process. Many open questions for these models such as heavy traffic for slow retrial case, waiting time distribution are still open for further investigations. Under multiserver settings, only the stability conditions are known~\cite{Morozov07,Morozov16c,Shin14,Dayar16}.  

\subsubsection{Constant retrial policy}
Multiclass M/G/1/1 retrial queues with constant retrial rates have been paid much attention in recent years. For this model, under Markovian assumptions, i.e., the service times of each class follow a distinct exponential distribution; the underlying Markov chain is multi-dimensional. In the case of two classes, boundary problems are formulated and some information on the means number of customers of two classes in the orbit may be derived. Furthermore, information on the joint generating functions of the numbers of customers in the orbit are obtained~\cite{Avrachenkov14} using Riemann-Hilbert boundary value problems. Song et al.~\cite{Song15} study the same model using a kernel approach and obtain tail asymptotic for the queue length of each class. Dimitriou~\cite{Dimitriou16a,Dimitriou16b} studies some extended models for network coding in relay nodes in wireless networks.

\subsection{Priority and Related Models}
\subsubsection{Classical retrial policy}
In retrial queues with priority, blocked priority customer can wait in front of the server while blocked normal customer joins the orbit. 
Under the classical retrial rate setting, bivariate generating functions for the joint queue length distribution of the number of customers in normal queue and that in the orbit is obtained in an integral form~\cite{Choi99,Fali97,Falin-Artalejo-Martin93}. For this model, heavy traffic analysis (the total traffic intensity tends to 1) is carried out by Falin et al.~\cite{Falin-Artalejo-Martin93}. Recently, Walraevens et al.~\cite{Joris_Dieter_Tuan} derive tail asymptotic formulas for the stationary distribution of the number of customers in the orbit. A detailed survey on related models is presented in~\cite{Choi99}.

\subsubsection{Constant retrial policy}
Priority retrial queues with constant retrial rate are also paid much attention. Lower priority customer joins the orbit while high priority customer joins the priority queue. The orbit operates as a single server queue where customers retry to occupy the server according to a FCFS manner and only the customer in the head of the orbit queue retries at a time. Under a pure Markovian setting, i.e., exponential retrial time and service time, Li and Zhao~\cite{Zhao05} obtain tail asymptotics of the priority queue given a fixed number of customers in the orbit queue. Gomez-Corral~\cite{Gomez99} considers the model with constant retrial rate where the retrial time of the customer in the head of the orbit and the service time are arbitrarily distributed. 
Atencia and Moreno~\cite{Atencia_Moreno} analyze the model under general retrial time and Bernoulli scheduling. 
In this model~\cite{Atencia_Moreno} an arriving customer that sees the server busy either joins the priority queue (normal queue) or the orbit queue. 
The authors~\cite{Atencia_Moreno} obtain the bivariate generating function of the number of customers in the orbit and that in the normal queue and the remaining service time or the remaining retrial time. The analysis of the models with both arbitrary service and retrial times is possible because at any time we need to keep track of either remain service time or remaining retrial time.

\subsection{Queueing Networks with Retrials}
In our everyday life, there are various service systems, in which customers are served by a number of servers in a certain order. For example, in a manufacturing system, a product is made and checked by a number of persons. In a computer system, a message is transmitted from a source to a destination through several devices such as computers, routers and switches. These systems can be modeled by queueing networks. 

Queueing networks with retrials do not possess product form solution~\cite{Artalejo_Economou05}. Thus, exact solution is obtained in a few special cases~\cite{phung5,Moutzoukis}. 
Phung-Duc~\cite{phung5} obtains explicit solution for a simple two-node tandem network with classical retrials at the first node. 
Moutzoukis and Langaris~\cite{Moutzoukis} derive the explicit results for the tandem model with constant retrial rate and blocking at the first server. For complex retrial queueing networks, a practical approach is the fixed point approximation. For example, fixed point approximations are used to analyze some tandem models with retrials by Avrachenkov et al.~\cite{Avrachenkov1,Avrachenkov2}. In fixed point approximation, 
the system is divided into multiple subsystems whose input parameters are unknown. Furthermore, these subsystems are assumed to be independent. The output of one subsystem is the input of another subsystem. After some iterative calculations, one will get a convergence determining all the unknown parameters. One drawback of this methodology is that a rigorous proof of the convergence and the accuracy of the approximation are not always presented. Moreover, the approximation is basically valid only under the slow retrial regime.
Numerical solutions for some simple tandem retrial models are presented in~\cite{Cskim10b,Cskim10,Taramin09,Gomez02}.

Recently, Fiems and Phung-Duc~\cite{Fiems16} present a light-traffic analysis for finite-source retrial systems arising from CSMA protocols without collision. These systems could be formulated by multidimensional Markov chains which are also used to represent retrial queueing networks. The analysis by Fiems and Phung-Duc~\cite{Fiems16} is based on series expansion subject to the arrival rate around the origin and is validated by simulation. Thus, power series expansion may be useful for analyzing queueing networks with retrials.

\subsection{Model with Orbital Search}
Orbital search mechanism for customers in the orbit is introduced by Artalejo et al.~\cite{Artalejo_Krishnamoorthy02}, where upon service completion, with some probability the server can pick a job from the orbit with zero searching time. This mechanism is further extended in~\cite{Dudin04,Deepak13,Krishnamoorthy05}.  Chakravarthy et al.~\cite{Chakravarthy06} investigate multiserver models where an idle server immediately picks up a customer from the orbit (zero searching time) with a probability or stays idle with the commentary probability. Recently, Dragieva and Phung-Duc~\cite{Phung-Duc-Dragieva,Dragieva16} propose a related model that the authors call the retrial queueing model with interaction between server and customers in the orbit. The main idea is that not only customers retry to capture an idle server (incoming calls) but the server also makes outgoing calls to retrial customers. The distributions of the durations of incoming calls and outgoing calls are different. Under M/M/1/1 settings, the authors in~\cite{Dragieva16} obtain explicit solution for the generating functions of the joint queue length distribution and the stability condition is obtained for the M/M/$c$/$c$ model.

\subsection{Game Theoretic Analysis}
Recently, game theoretic analysis of queues has been attracted much attention. Some authors study game theoretic analysis for retrial models. In particular, a series of works by Economou and Kanta~\cite{Economou08,Economou11} provide detailed analysis for retrial models with constant retrial rate. Wang et al.~\cite{Wang-Zhang13,Wang-Wang16,Wang-Li16} analyze also the models with classical retrial rate. To be more precise, models in Wang et al.~\cite{Wang-Wang16,Wang-Li16} are devoted to strategic joining behavior of secondary users in cognitive networks. It should be noted that these references are devoted to M/M/1/1 type retrial queues only. Thus, the analyses of more general models might be promising future topics.  

\subsection{General Retrial Times}
Most of researches assume the exponential retrial time. Only a few references are devoted to models with other retrial time distributions where each customer in the orbit acts independently of other customers~\cite{Chakravarthy13,Shin11,Shin14a}. In the current literature, there is not an exact analysis for this type of settings and only some approximations or simulations are presented. Thus, researches in this direction may be highly appreciated.  

\subsection{Other Performance Measures}
In almost the work mentioned above, the main quantity of interest is the stationary queue length distribution. 
A few references are devoted to some new quantities of interest. In particular, channel idle period is analyzed by Artalejo and Gomez-Corral in~\cite{Artalejo-Gomez03}. Distributions of the successful and blocked events are studied in~\cite{Amador-Artalejo07,Amador-Artalejo09a,Amador-Artalejo09b}. Maximum queue length in busy period is presented in~\cite{Artalejo-Economou-Lopez} while that in a fixed time interval is analyzed by Gomez-Corral and Garcia~\cite{Gomez-Garcia}. Artalejo and Lopez-Herrero~\cite{Artalejo-Lopez07} analyze the distribution of the number of retrials of a tagged customer in M/G/1/1 and M/M/$c$/$c$ retrial queues. Falin~\cite{Fali86} obtain the distribution of the number of retrials for M/G/1/1 retrial queues. Dragieva~\cite{Dragieva13,Dragieva14} studies the distribution of the number of retrials in model with finite source with arbitrary service time distribution. Gao et al.~\cite{Gao16} obtain the distribution for number of retrials in an M/M/1/1 retrial queue with constant retrial rate and impatient customers.

\section{Concluding Remarks}
In this paper, we have surveyed the main theoretical results for retrial queueing models. We have also investigated retrial queueing models arising from real applications such as call centers, random access protocols, cellular networks etc. We hope that this paper can be served as a basic reference for researchers who want to enter and deepen this field. Because the retrial queue literature is rich, we also refer to some earlier survey papers~\cite{Yang87,Fali90,Artalejo99,Gome06,Artalejo10}, two books~\cite{Fali97,artalejo08} and the recent Special Issue~\cite{Gomez-Phung}. Most of references in this paper is for continuous time retrial queues. We refer to Nobel~\cite{Nobel16} for a survey on results of discrete time retrial queues. Sections 1 and 2 are partially based on the dissertation of the author~\cite{Phung-dissertation}.

\section*{Acknowledgments}
Tuan Phung-Duc was supported in part by JSPS KAKENHI Grant Number 26730011. The author would like to thank two anonymous reviewers whose comments greatly helped to improve the presentation of the paper. I would like to thank the Editors of the book, especially Professor Shoji Kasahara for giving me an opportunity to write this survey.

The author would like to devote this paper to the memory of Professor Jesus Artalejo who was a leading researcher in the fields of Queueing Theory and Mathematical Biology publishing more than one hundred papers and was a co-author and friend of the author of this paper.



\end{document}